\documentclass[showpacs, pra,twocolumn,preprintnumbers ,amsmath, amssymb, superscriptaddress, aps]{revtex4-2}

\usepackage{color}
\usepackage{amsmath,amssymb}
\usepackage[mathscr]{euscript}
\usepackage{pifont}
\usepackage{amssymb}  
\usepackage{bbold}
\usepackage{float}
\usepackage{subfloat}

\usepackage[caption=false]{subfig}
\usepackage{tikz}
\usepackage{makecell}
\usepackage{subfig}
\usepackage{pifont}   
\usepackage{graphicx} 
\usepackage{dcolumn}  
\usepackage{bm}       
\usepackage{multirow} 
\usepackage{placeins}
\usepackage[colorlinks]{hyperref}
\usepackage{mathtools}

\newcommand{\eqfitpage}[1]{\resizebox{\linewidth}{!}{$#1$}}

\begin{document}

\title{Band structure and band topology in twisted homotrilayer transition metal dichalcogenides}
\author{Hassan AlBuhairan}
 \email{hassan.albuhairan@kfupm.edu.sa}
		\affiliation{Department of Physics, King Fahd University of Petroleum and Minerals, 31261 Dhahran, Saudi Arabia}
\author{Michael Vogl}
 \email{ssss133@googlemail.com}
	\affiliation{Department of Physics, King Fahd University of Petroleum and Minerals, 31261 Dhahran, Saudi Arabia}
\begin{abstract}
    We investigate a simplified continuum model of a twisted homotrilayer TMD with negligible next-nearest layer couplings. We systematically analyze band structure and topology of various stacking configurations in a twist angle range from $1^\circ$ to $4^\circ$. This allows us to uncover a plethora of topological transitions as well as various angle regimes with flat bands - some of them topologically non-trivial, which are of growing interest for the realization of exotic strongly correlated phases. Additionally, we uncover surprising properties, such as that for certain stacking configurations some layers effectively decouple from other layers. In this case the only remnant of coupling between layers is a layer-dependent energy shift.
\end{abstract}
\maketitle

\section{Introduction}

Moir\'e superlattices are long-wavelength periodic modulations that form when two adjacent layers of a lattice have slight differences in lattice constants or have the same lattice constant but are rotated
with respect to one another. Such moir\'e systems often produce spatial confinement, resulting in flat bands in which electron-electron or electron-phonon interactions dominate the energy scale. Therefore, electrons in moir\'e superlattices often exhibit strongly correlated phases. Experiments involving these flat bands have led to many exciting discoveries, such as unconventional superconductivity in twisted bilayer graphene (TBG) \cite{Cao2018}, which by itself has led to a surge of interest in twisted two-dimensional materials. Besides superconducting phases \cite{Cao2018, Lu2019, Yankowitz2019,Stepanov2020,Saito2020}, this material class also gives rise to other exotic quantum phases like correlated insulating phases \cite{Lu2019, Sharpe2019, Serlin2020} and orbital magnetism \cite{Cao2018a,Lu2019, Zondiner2020, Wong2020}, making it one of the currently most exciting research areas in condensed matter theory. In addition to these interacting phases, TBG  was also a subject of extensive studies on non-interacting properties such as band structure and topological phases both in the equilibrium case \cite{Gail2011, Song2019, Nuckolls2020} and in a non-equilibrium setting \cite{Li_2020_twist_floquet,PhysRevResearch.1.023031,RODRIGUEZVEGA2021168434,PhysRevB.101.235411,PhysRevB.101.241408,Lu_2021}. Besides TBG, similarly interesting phases both in the interacting and non-interacting regimes have been discovered in other graphene-based twisted materials, including double-bilayer graphene \cite{Liu2020,Cao2020,Shen2020,Burg2019,PhysRevResearch.2.033494} and trilayer graphene \cite{Chen2019,Chen2019a,Chen2020,PhysRevB.104.195429}. 

In addition to graphene, transition metal dichalcogenides (TMDs) emerged as a promising material that shares some of its features. However, unlike graphene, it has a large direct bandgap \cite{mak2010, Lebegue2009, Li2007, Splendiani2010}, which leads to its own distinct set of interesting phenomena. To further contrast graphene, TMDs exhibit strong spin-orbit splitting and the interesting phenomenon of spin-valley locking, where spin degrees of freedom become linked to the momentum degree of freedom - specifically, spins become associated with the $K$ valleys at the corners of the Brillouin zone. These phenomena mean that twisted TMDs, like TBG, form unique platforms for exciting physics. For instance, they play host to various idealized models of physics such as a Hubbard model on a triangular lattice \cite{Wu2018, Wu2019, Tang2020, MoralesDuran2022}, generalized Kane-Mele models \cite{Wu2019, Devakul2021} and generalized Wigner crystals \cite{Regan2020, Padhi2021}. They also are predicted to play host to a highly-tunable phase diagram of a variety of magnetic phases \cite{PhysRevB.104.214403,Vogl_2022_light_magnetic} as well as an extended Hubbard model with highly non-local interactions \cite{PhysRevLett.128.217202}. Much like TBG, twisted TMDs are not only exciting in the interacting limit but also host interesting non-interacting topological phases \cite{Wu2019,Vogl_2021_floquet_homobilayerTMD}.

Theoretical investigations of twisted TMDs were conducted using various methods ranging from atomistic ab initio approaches like density functional theory (DFT) to effective continuum models. For instance, using DFT and neglecting the effects of spin-orbit coupling, the band structure of MoS$_2$, a two-dimensional TMD, was analyzed and shown to exhibit flat valence bands with widths of 23 and 5 meV at twist angles of  $\theta=3.5^\circ$ and $\theta=56.5^\circ$, respectively. A similar analysis was performed for various other TMD homobilayers \cite{Naik2018}. However, using DFT is computationally expensive due to its scaling with the system size, which makes smaller angles calculations increasingly more difficult.

As a computationally cheaper but still atomistic alternative, one may employ tight-binding Hamiltonians. For instance, tight-binding Hamiltonians exist for untwisted homobilayer TMDs as proposed in Ref.\ \cite{Fang2015}. Such a tight-binding approach has the advantage that it permits including spin-orbit couplings in a computationally cheap fashion. Furthermore, such models permit reaching smaller twist angle regimes than it would be computationally practical using DFT. Based on this model, band structures of homobilayers MoS$_2$ were studied in the small twist angle regime. Particularly, widths of flat bands and bandgaps were studied as a function of twist angle and other structural parameters \cite{Zhan2020, Zhang2020}. A similar analysis was also performed for a variety of other twisted homo- and hetrobilayer TMDs \cite{Vitale2021}. 

Another alternative theoretical approach to the aforementioned atomistic methods is to use continuum effective mass models \cite{Jung2014, Wu2018, Wu2019}, which, despite being limited to small angles and relatively low energies, are computationally very efficient and thus advantageous for systematic studies. In particular, the continuum model for twisted TMD homobilayers proposed in Ref.\ \cite{Wu2019} was used in further theoretical investigations of this attractive material class. For example, a systematic study of single-particle and many-body moir\'e physics in twisted WSe$_2$, including a complete topological phase diagram that studies the effects of introducing a layer potential difference, was presented in Ref.\ \cite{Pan2020}. Furthermore, a combination of the continuum model with other atomistic methods, such as DFT, was applied to twisted TMDs homobilayers (with WSe$_2$ as a focus of the work). Here, a particular magic twist angle of $\theta=1.43^\circ$ with almost perfectly flat bands was identified, and the realization of several exotic physical models, such as an interaction-driven Haldane insulator, was predicted \cite{Devakul2021}.

In this work, we expand on the theoretical efforts of investigating twisted TMDs by systematically studying the electronic and topological properties of various twisted homotrilayer TMD configurations using a simplified continuum model that neglects next nearest layer couplings. We observe the existence of flat bands in all configurations, especially at small twist angles, which sometimes remain flat for a wide range of twist angles $\theta>1^\circ$. The bands in our model exhibit other interesting properties. For example, in the middle-twist configurations at a twist angle $\theta=1.2^\circ$, we observe that the top three valence bands are all extremely flat, with bandwidths as low as $\approx 0.013$ meV, where the topmost band is separated from the rest by a comparatively huge bandgap of $24$ meV. The second and third bands are in very close proximity to one another, and both are topologically non-trivial. This work demonstrates some of the fascinating features of trilayer TMDs and highlights their potential as an exciting subject of future research on moir\'e materials. 

This paper is organized as follows. In Sec.\ \ref{sec:model}, we discuss our model Hamiltonian, define the different stacking configurations of the moir\'e system, explain the method used to obtain band structures, and outline the method employed to compute valley Chern numbers. In Sec.\ \ref{sec:results}, we analyze band structures, highlight twist angles with flat bands - magic angles - and identify band touching points. We further identify topological phase transitions and non-trivial topological phases. Our findings are supported by plots of the valley Chern numbers, bandwidths, and bandgaps of the topmost valence bands as a function of twist angle taking values in the range $1^\circ \leq \theta \leq 4^\circ$. Finally, in Sec.\ \ref{sec:conclusion}, we conclude by giving a summary and an outlook on our main findings.

\section{Physical system and model}\label{sec:model}

In this section we will lay out the theoretical foundation for our study of twisted homotrilayer TMDs. The Hamiltonian we will study can be applied to the study of various TMDs, but we will specifically focus on parameters that are appropriate for MoTe$_2$ to produce concrete numerical results. MoTe$_2$ has a hexagonal lattice structure with a lattice constant $a_0=3.472$ \AA.  More preciselu it consists of two layer of Te atoms with a layer of Mo atoms between them. For our study of trilayer MoTe$_2$ build on the bilayer model in Ref.\ \cite{Wu2019} and extend it to a description of three layers. Hereby, we assume that hoppings between non-adjacent TMD layers can be neglected because tight binding hoppings typically decay exponentially with distance. In our model we focus on the valence band maxima that are located in $\pm K$ valleys. Since the two valleys are related by time reversal symmetry, we choose to focus on the $+K$ valley. Valence band states, here, are subject to relatively strong spin-orbit splitting. This allows us to choose a description that focuses solely on the  spin-up states, which are closer to the Fermi energy. According to this logic the two band Hamiltonian for two layers from Ref.\ \cite{Wu2019}, in our case of three layers, generalizes to a three-band $k.p$ Hamiltonian:

\begin{equation}
\eqfitpage{
\begin{aligned}
    &\mathcal{H}_0(\mathbf{k},\boldsymbol{\delta_1},\boldsymbol{\delta_3})=\\&\left(
\begin{array}{ccc}
 \frac{-\hbar^2 \mathbf{k}^2}{2m^*}+\Delta_1(\boldsymbol{\delta_1}) & \Delta^T_{12}(\boldsymbol{\delta_1}) & 0 \\
  \Delta^T_{12}(\boldsymbol{\delta_1})^\dagger &  \frac{-\hbar^2 \mathbf{k}^2}{2m^*} +\Delta_{2a}(\boldsymbol{\delta_1})+\Delta_{2b}(\boldsymbol{\delta_3}) &  \Delta^T_{23}(\boldsymbol{\delta_3}) \\
 0 &   \Delta^T_{23}(\boldsymbol{\delta_3})^\dagger &   \frac{-\hbar^2 \mathbf{k}^2}{2m^*}+\Delta_3(\boldsymbol{\delta_3})
\end{array}
\right)
\end{aligned}}
\end{equation} where $\mathbf{k}$ is the momentum measured from the $+K$ point, $m^*=0.62 m_e$ is the valence band effective mass, $\boldsymbol{\delta}_i$ is a displacement vector that captures the relative orientation between layer $i$ and layer 2 (we take layer 2 as a reference). Furthemore,  $\Delta_i$ is a potential energy pertaining to layer $i$, and $\Delta^T_{ij}$ is an interlayer tunneling amplitude between layers $i$ and $j$. Both atoms from layer 1 ($\Delta_{2a}$) and layer 3 ($\Delta_{2b}$) contribute to the effective potential $\Delta_2=\Delta_{2a}+\Delta_{2b}$ on layer 2. We stress that we neglect interlayer hopping between layers 1 and 3. Expressions for $\Delta_i$ and  $\Delta^T_{ij}$ can be obtained in lowest harmonic approximation by observing how their dependence on $\boldsymbol{\delta_i}$ is constrained by the symmetry properties \cite{Wu2019}, which yields 

\begin{equation}
\Delta_i(\boldsymbol{\delta_i})=
        2V \sum_{j=1,3,5} \cos({\mathbf{G}_j.\boldsymbol{\delta_i}+s \psi})
\end{equation}
\begin{equation}
\Delta^T_{ij}(\boldsymbol{\delta_i})=
        w(1+e^{-i \mathbf{G}_2.\boldsymbol{\delta_i}}+e^{-i \mathbf{G}_3.\boldsymbol{\delta_i}})
\end{equation} where $V=8$ meV characterizes the amplitude of the potential, $\mathbf{G}_j$ is a reciprocal lattice vector obtained by the counterclockwise rotation of $\mathbf{G}_1=(4\pi)/(\sqrt{3}a_0)\hat{y}$ by angle $(j-1)\pi/3$, $s=1$ for $\Delta_1$ and $\Delta_{2b}$, $s=-1$ for $\Delta_{2a}$ and $\Delta_{3}$, $\psi = -89.6 ^{\circ}$ characterizes the shape of the potential, and $w = -8.5$ meV is a tunneling strength parameter.

In this work, we study different configurations of trilayer MoTe$_2$ that are constructed from combinations of layer twists and shifts. To have a fixed coordinate system, we take the center of layer 2 as the origin.

There are two types of twists that can be considered: i) a top twist, achieved by twisting layer 1 by an angle $+\theta/2$ and both layers 2 and 3 by an angle $-\theta/2$, and ii) a middle twist, achieved by twisting layer 2 by an angle $-\theta/2$ and both layers 1 and 3 by an angle $+\theta/2$, where all twists are taken around a rotation axis that passes through the origin. Note that a bottom twist would be equivalent to a top twist, which is why we do not consider this case. For layer shifts, we restrict our investigation to shifting layer 3 by the high-symmetry displacements $n(\mathbf{a}_1+\mathbf{a}_2)/3$ for $n=0, \pm 1$, where $\mathbf{a}_1=a_0(1,0)$ and $\mathbf{a}_2=a_0(1/2, \sqrt{3}/2)$ are the primitive lattice vectors. Hence, we consider six families of configurations of trilayer MoTe$_2$, which we denote $C_{t,m}^{n=0,\pm1}$, where the subscript refers to the type of twist and the superscript to the high-symmetry displacement index of layer 3. The twist and shift operations are described by the parameters $\boldsymbol{\delta}_{1}=\mathbf{d}_{1}-\mathbf{d}_2$ and $\boldsymbol{\delta}_{3}^n=\mathbf{d}_{2}-\mathbf{d}_3+\boldsymbol{\delta}_{0}^n$, where $\mathbf{d_i}=\theta_i \hat{z} \times \mathbf{r}$, $\theta_i$ is the twist angle of layer $i$, $\mathbf{r}$ is the spatial position, and $\boldsymbol{\delta}_{0}^n=n(\mathbf{a}_1+\mathbf{a}_2)/3$.

\begin{table}[H]
    \centering
    \begin{tabular}{|c|c|c|}\hline
         Configuaration&$\boldsymbol{ \delta}_1$&$\boldsymbol{ \delta}_3^n$ \\\hline
         $C_{t}^{0}$ & $\theta\hat{z} \times \mathbf{r}$&0\\\hline
         $C_{t}^{-1}$& $\theta\hat{z} \times \mathbf{r}$&$-(\mathbf{a}_1+\mathbf{a}_2)/3$\\\hline
         $C_{t}^{1}$& $\theta\hat{z} \times \mathbf{r}$&$(\mathbf{a}_1+\mathbf{a}_2)/3$\\\hline
         $C_{m}^{0}$& $\theta\hat{z} \times \mathbf{r}$&$-\theta\hat{z} \times \mathbf{r}$\\\hline
         $C_{m}^{-1}$& $\theta\hat{z} \times \mathbf{r}$&$-\theta\hat{z} \times \mathbf{r}-(\mathbf{a}_1+\mathbf{a}_2)/3$\\\hline
         $C_{m}^{1}$ & $\theta\hat{z} \times \mathbf{r}$&$-\theta\hat{z} \times \mathbf{r}+(\mathbf{a}_1+\mathbf{a}_2)/3$\\\hline
    \end{tabular}
    \caption{Summary of different twist and shift configurations}
    \label{tab:my_label}
\end{table}

\begin{figure}[t!]
\centering\graphicspath{{./Figures/}}
\includegraphics[width=\linewidth,clip]{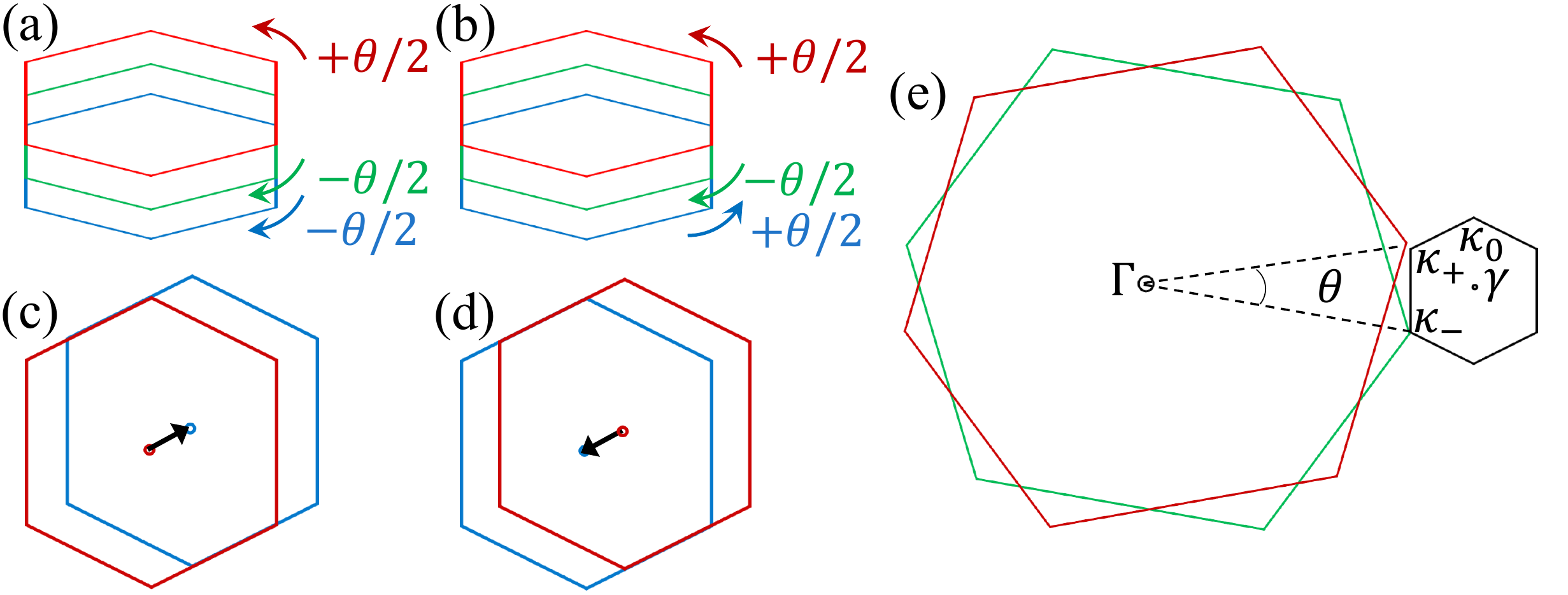}
\caption{(a) A top-twist of trilayer MoTe$_2$ is realized by twisting layer 1 (red) by an angle $+\theta/2$ and both layers 2 and 3 (green and blue) by an angle $-\theta/2$, whereas a middle-twist (b) is realized by twisting layer 2 by an angle $-\theta/2$ and both layers 1 and 3 by an angle $+\theta/2$. We consider shifting layer 3 by a high-symmetry displacement $n(\mathbf{a}_1+\mathbf{a}_2)/3$, a top view of $n=1$ and $n=-1$ shifts are shown in (c) and (d), respectively. (e) Brillouin zones of layer 1 (red) and layer 2 (green) for top-twist configuration (the Brillouin zone of layer 3 is hidden), and the moir\'e Brillouin zone (black).}\label{fig.config}
\end{figure}

\begin{figure*}[t!]
\centering\graphicspath{{./Figures/}}
\includegraphics[width=\linewidth]{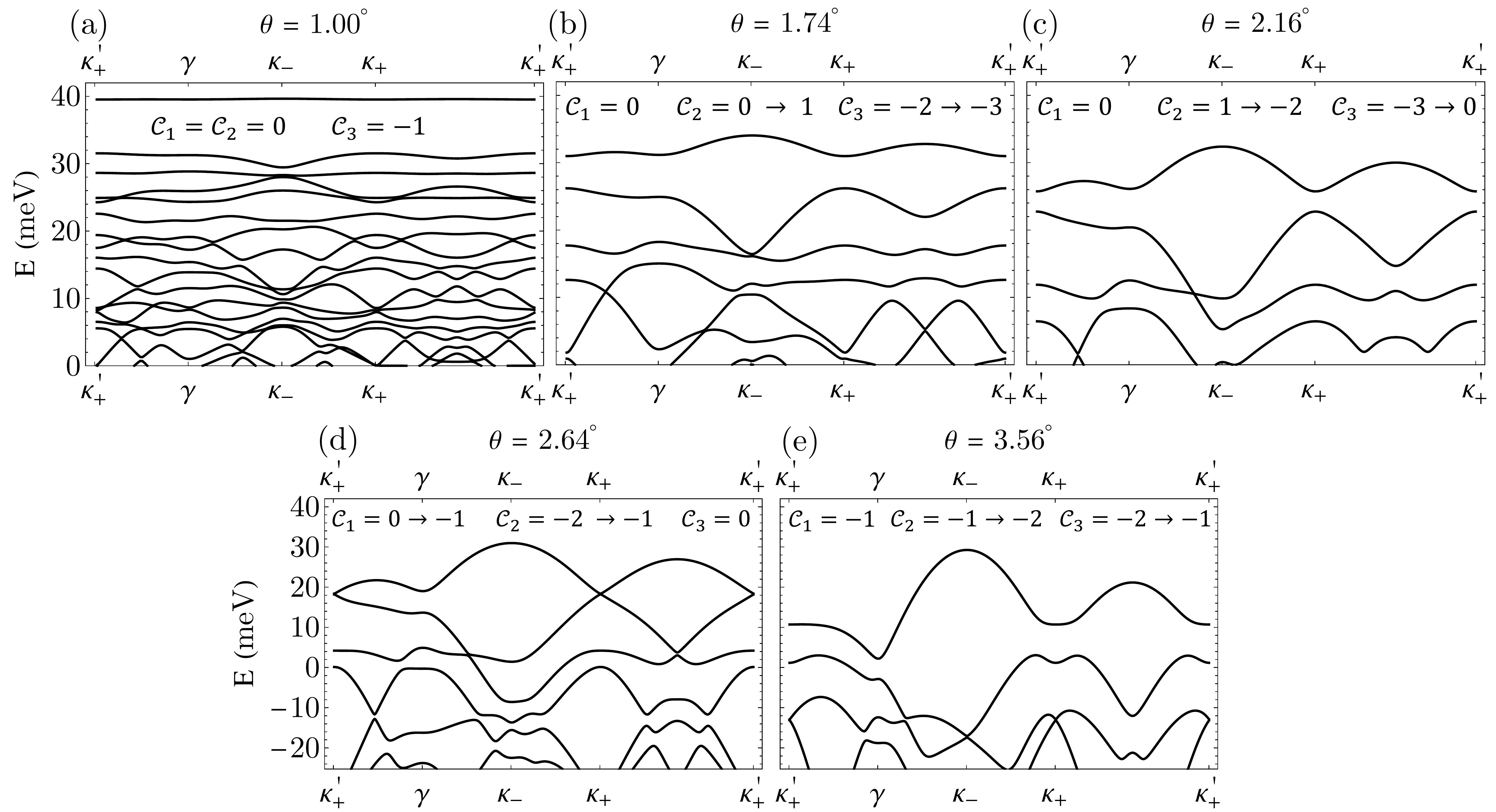}
\caption{Moir\'e band structures of $C_t^0$ for twist angles (a) $\theta=1.00^\circ$, (b) $\theta=1.74^\circ$, (c) $\theta=2.16^\circ$, (d) $\theta=2.64^\circ$, and (e) $\theta=3.56^\circ$. $\mathcal{C}_i$ denotes the Chern number of the $i$-th band (counting from above). }\label{fig:Ct0}
\end{figure*}

The moir\'e Hamiltonian for the configurations with twists is given by

\begin{equation}
\label{eq:moire-hamiltonian}
\eqfitpage{
\begin{aligned}
    &\mathcal{H}(\mathbf{k},\mathbf{r},\boldsymbol{\delta}_1,\boldsymbol{\delta}_3^{n})=\\&\left(
\begin{array}{ccc}
 \frac{-\hbar^2 (\mathbf{k}-\boldsymbol{\kappa_+})^2}{2m^*}+\Delta_1(\boldsymbol{\delta}_1) & \Delta^T_{12}(\boldsymbol{\delta}_1) & 0 \\
  \Delta^T_{12}(\boldsymbol{\delta}_1)^\dagger &  \frac{-\hbar^2(\mathbf{k}-\boldsymbol{\kappa_-})^2}{2m^*} +\Delta_{2a}(\boldsymbol{\delta}_1)+\Delta_{2b}(\boldsymbol{\delta}_3^{n}) &  \Delta^T_{23}(\boldsymbol{\delta}_3^{n}) \\
 0 &   \Delta^T_{23}(\boldsymbol{\delta}_3^{n})^\dagger &   \frac{-\hbar^2 (\mathbf{k}-\boldsymbol{\kappa_\pm})^2}{2m^*}+\Delta_3(\boldsymbol{\delta}_3^{n})
\end{array}
\right)
\end{aligned}}
\end{equation} where $\mathbf{r}$ is the position vector, $\boldsymbol{\kappa_\pm}$ are vectors that are located at the corners of the moir\'e Brillouin zone, as illustrated in Fig.\ \ref{fig.config}. In the last element of $\mathcal{H}$, $\boldsymbol{\kappa}_-$ corresponds to a top twist and  $\boldsymbol{\kappa}_+$ to a middle twist. Note that shifting the momenta by $\boldsymbol{\kappa_\pm}$ allows us to maintain a consistent definition of $\mathbf{k}$, i.e., consistent choice of coordinate system, for all layers despite the various rotations involved.

To investigate the electronic properties, we numerically calculate moir\'e band structures near the $+K$ point, see Fig.\ \ref{fig.config}(e). These are obtained by diagonalizing the moir\'e Hamiltonian in Eq.\ \eqref{eq:moire-hamiltonian} using a plane-wave expansion based on Bloch's theorem,

\begin{equation}
 \mathcal{H}_{nm}(\mathbf{q})=\left\langle \phi_{n} \right\vert   \mathcal{H}(\hat{\mathbf{k}}+\mathbf{q})  \left\vert \phi_{m } \right\rangle,
\end{equation}
where $\phi_{n }$ is a plane wave state corresponding to a reciprocal lattice vector with label $n$ and $q$ is the crystal momentum. Bloch's theorem was used to cast the Hamiltonian into this form that allows the expansion in terms of plane waves. The set of reciprocal lattice vectors that is used is truncated in a way that respects sixfold rotational symmetry to avoid truncation artifacts.

To understand topological features we compute the Chern numbers by discretizing the Brillouin zone using the Fukui-Hatsugai-Suzuki method \cite{Fukui2005}, which we summarize below. We start by discretizing the Brillouin zone (in our case as a $30\times30$ grid, which was enough points to ensure convergence of the Chern numbers in all cases considered in this work) and denote its reciprocal lattice points by $\mathbf{k}_l$, where $l$ is an integer that spans the lattice positions. We then define so-called link variables,
\begin{equation}
    U_{\alpha,n}(\mathbf{k}_l)= 
    \frac{\left\langle \phi_{n}(\mathbf{k}_l)|\phi_{n}(\mathbf{k}_l+\boldsymbol \mu_{\alpha}) \right\rangle}{|\left\langle \phi_{n}(\mathbf{k}_l)|\phi_{n}(\mathbf{k}_l+\boldsymbol \mu_{\alpha}) \right\rangle|},
    \label{eq:U}
\end{equation} here $\boldsymbol\mu_{\alpha}=\hat{\alpha} L_{\alpha}/N$, where $\alpha$ refers to the $x$ or $y$ coordinate, $L_{\alpha}$ is the length of the lattice in the $\alpha$ direction, and $N$ is the number of points (30 points in our case). We then define the lattice field strength at site $l$ and for band $n$ as
\begin{equation}
    F_n(\mathbf{k}_l)=
    \text{ln} \left( \frac{U_{x,n}(\mathbf{k}_l) U_{y,n}(\mathbf{k}_l+ \boldsymbol \mu_x)}{ U_{x,n}(\mathbf{k}_l+\boldsymbol \mu_y)^{-1} U_{y,n}(\mathbf{k}_l)^{-1}} \right)
    \label{eq:field-strength}.
\end{equation} The valley Chern number of the $n$-th band is then given by
\begin{equation}
   \mathcal{C}_n= \frac{1}{2\pi i}\sum_{l} F_n(\mathbf{k}_l).
\end{equation} In the analysis that follows we focus on Chern numbers $\mathcal{C}_n$ for the top three valence bands, which we label in a descending order of energy, i.e., $\mathcal{C}_n$ is the Chern number of the $n$-th band from the top.

As possible twist angles, we consider $\theta \in [1.0^\circ, 4.0 ^\circ]$. The lower bound is set such that a numerically reasonable number of plane waves is needed to get convergence for the band structure and Chern numbers in the chosen energy range. For this we use 91 $K$ points (number of plane waves), which were chosen in such a way that our choice preserves the sixfold rotational symmetry. The upper limit of $4.0 ^\circ$ is set to ensure that we stay in an angle range where a twist operation can be safely treated as linear in angle, i.e., $\delta_i\approx \theta \hat z\times \vec r$ (see Tab.\ \ref{tab:my_label}). Going further beyond this angle range would require a treatment involving rotation matrices, which would introduce a second length scale into the Hamiltonian. Such a second length scale gives rise to a quasicrystal structure when it is non-commensurate (i.e. a non-rational fraction) with the first length scale. In such a case Bloch's theorem would not be applicable anymore. Of course, this is a very exciting regime but not the topic of this work but has attracted recent attention \cite{doi:10.1021/acs.nanolett.0c00172,doi:10.1021/acsnano.9b07091,doi:10.1073/pnas.1720865115,PhysRevB.99.165430,Yu2019quasicrystal}.

\section{Results and Discussion}\label{sec:results}

\begin{figure*}[t!]
\centering\graphicspath{{./Figures/}}
\includegraphics[width=0.75\linewidth]{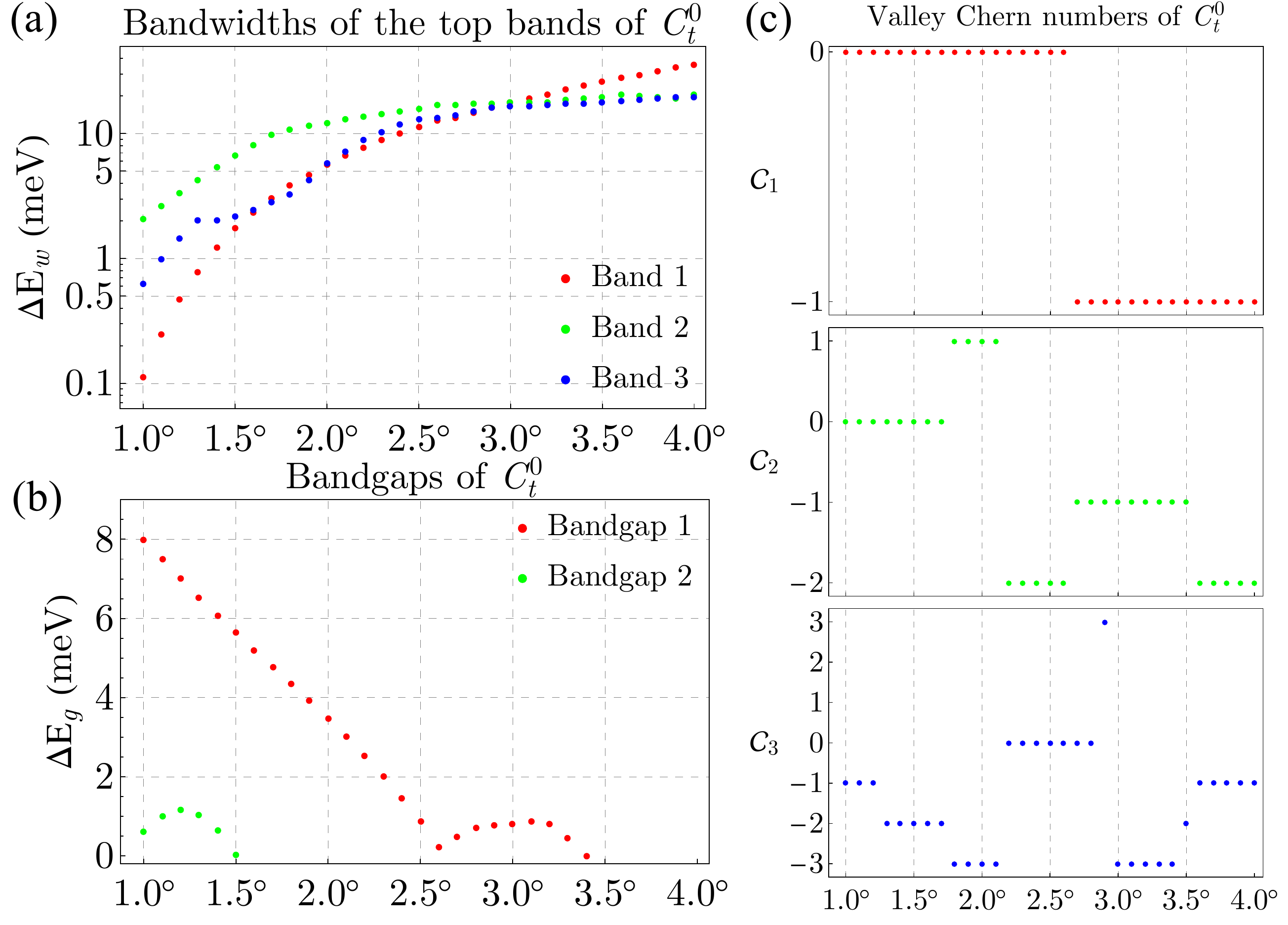}
\caption{Summary of $C_t^0$ calculations. (a) Bandwidths of the top three bands in a logarithmic scale of energy at different twist angles $\theta$. (b) The first two bandgaps as a function of $\theta$, where bandgap 1 refers to the gap between band 1 and band 2, and bandgap 2 refers to the gap between band 2 and band 3. (c) Valley Chern numbers of the top three bands as a function of $\theta$.}\label{fig:Ct0-summary}
\end{figure*}

In this section, we discuss the most interesting electronic and topological features of the different stacking configurations of the twisted homotrilayer TMD as a function of twist angle $\theta$. We base our discussion for each configuration on two themes of results presented in separate figures: (i) band structure calculations, in which we show the flat bands and capture band touchings with the second band (we discuss but do not show touchings between the third and lower bands as they happen frequently), and (ii) summarized numerical calculations, in which we show the bandwidths, bandgaps, and valley Chern numbers of the top three bands as a function of $\theta$. 

\subsection{Top twist with no relative shift between the bottom two layers $C_t^0$} 
\label{sec:toptwst_noshift}
As first configuration, we consider $C_{t}^{0}$ (see Tab.\ \ref{tab:my_label}), which is a trilayer where the top layer is rotated with respect to the bottom two layers and there is no relative shift between the bottom two layers -they are exactly on top of one another. In what follows below we discuss all the physical effects we observe as twist angles are increased from an angle $1^\circ$ up to the final value of $4^\circ$.

In Fig.\ \ref{fig:Ct0}(a), which shows the band structure of $C_t^0$ at $\theta=1.00^\circ$, we observe extreme flat bands. Indeed, the first band (at $E \approx 40$ meV) is extremely flat with with bandwidth $\Delta E_{w1} \approx 0.1$ meV and so is the third band (at $E \approx 29$ meV) with a bandwidth of $\Delta E_{w3} \approx 0.6$ meV (see Fig.\ \ref{fig:Ct0-summary}(a)). Here, the first two bands are topologically trivial while the third band is non-trivial with a valley Chern number $\mathcal{C}_3=-1$. At this angle, $\theta=1.00^\circ$, the bandgap between the first and second bands  - we call it bandgap 1 - is at its maximum value (within our chosen angle range) of $\Delta E_{g1}=8$ meV, see Fig.\ \ref{fig:Ct0-summary}(b), whereas the bandgap between the second and third bands  - we call it bandgap 2 - is only $\Delta E_{g2} \approx 0.5$ meV. Such a bandgap $\Delta E_{g1}$ that is two orders of magnitude larger than the bandwidth $W$ of the first band is interesting because it suggests that there is a frequency $\omega$ regime $W<\omega<\Delta$ where one could periodically drive the system without having to worry about resonances - for example via light of frequency $\omega$ (in this case a safe frequency range would be in the infra red regime). A second reason the flat bands like the ones observed are interesting is because they can often be found to play host to strongly correlated phases. Indeed, flat bands contribute little as a kinetic term, which makes interactions between electrons relatively more important - an often important condition for the existence of strongly correlated phases.

In Fig.\ \ref{fig:Ct0-summary}(b) we see that as we increase the twist angle $\theta$, bandgap 1 starts to decrease linearly. To contrast this gap 2 first increases slightly up to $\theta \approx 1.2^\circ$, where it reaches its maximum of $E \approx 1$ meV. Next in Fig.\ \ref{fig:Ct0-summary}(c), at angle $\theta \approx 1.29^\circ$ we observe that the valley Chern number of the third band changes  $\mathcal{C}_3=-1\xrightarrow{}-2$ as it touches lower bands. We observe that the first and third bands have a bandwidth less than $\Delta E_w \approx 2$ meV for a substantial angle range up to $\theta=1.5^\circ$ (Fig.\ \ref{fig:Ct0-summary}(a)). While there seems to be no consensus in the literature on what is considered a flat band (some defining them as bands with small derivatives and some as bands with a width smaller than some cut-off), we state that for this angle range bands 1 and 3 can be considered as relatively flat - at least when compared to the initial gap between bands 1 and 2. 


The next interesting observation occurs at $\theta \approx 1.74^\circ$ (see Fig.\ \ref{fig:Ct0}(b)), where the second band crosses the third band at $\mathbf{k}=\boldsymbol \kappa_-$, resulting in a change in valley Chern numbers for both bands as $\mathcal{C}_2=0\xrightarrow{}1$ and $\mathcal{C}_3=-2\xrightarrow{}-3$. Therefore, the gap between the first trivial band and the second non-trivial band is expected to play host to an edge mode. Notice that bandgap 2 in this case is not zero, as we see in Fig.\ \ref{fig:Ct0-summary}(b), but $\Delta E_{g2} \approx -2.5$ meV, as the top of band 3 is not located at the touching point. We observe a kink at this angle in Fig.\ \ref{fig:Ct0-summary}(b) that correlates with this topological transition.  

Interestingly, another crossing between the second band and third band occurs at $\theta=2.16 ^\circ$. This band crossing happens at a $\mathbf{k}$-point along the $\boldsymbol \gamma -\boldsymbol \kappa_-$ path, see Fig.\ \ref{fig:Ct0}(c), which results in the valley Chern numbers changing as $\mathcal{C}_2=1\xrightarrow{}-2$ and $\mathcal{C}_3=-3\xrightarrow{}0$.

Next, at $\theta=2.64^\circ$, we observe that the first band and the second band touch at $\mathbf{k}=\boldsymbol{\kappa_+}$ and $\mathbf{k}=\boldsymbol{\kappa_+'}$ (see Fig.\ \ref{fig:Ct0}(d)). This band touching is accompanied with a change of Chern numbers. Indeed, the upper most band for the first time becomes topologically non-trivial as $\mathcal{C}_1=0\xrightarrow{}-1$ and $\mathcal{C}_2=-2\xrightarrow{}-1$. One may notice that the simultaneous crossing at the two k-points $\boldsymbol\kappa_+$ and $\boldsymbol\kappa_+'$ is a manifestation of inversion symmetry in the k-space, which is not broken by a change in twist angle. Furthermore, at the same angle, $\theta=2.64^\circ$, we observe in Fig.\ \ref{fig:Ct0-summary}(b) that the bandgap 1 displays a local minimum of zero gap size accompanied with a discontinuity. The discontinuity in the gap size correlates with the topological transition in this case.

Various additional topological phase transitions are identified for the third band and are displayed in Fig.\ \ref{fig:Ct0-summary}(c). An interesting such transition occurs at $\theta \approx 2.9^\circ$. Here, the third band's topology undergoes a series of sudden changes as $\mathcal{C}_3=0\xrightarrow{}3\xrightarrow{}-3$, which happens due to multiple crossings with lower bands (such as band 4) that occur at intermediate angles close to $\theta \approx 2.9^\circ$. Such crossings between the third band and lower bands happens again at $\theta \approx 3.49^\circ$, resulting in a change in valley Chern number $\mathcal{C}_3=-3\xrightarrow{}-2$.

For the angle range we consider, the last exciting change happens at $\theta \approx 3.56^\circ$, where the second band and the third band touch at $\mathbf{k}=\boldsymbol\kappa_-$, see Fig.\ \ref{fig:Ct0}(e). This band crossing is accompanied by a change in valley Chern numbers according to $\mathcal{C}_2=-1\xrightarrow{}-2$ and $\mathcal{C}_3=-2\xrightarrow{}-1$. 

\begin{figure}[t!]
\centering\graphicspath{{./Figures/}}
\includegraphics[width=\linewidth]{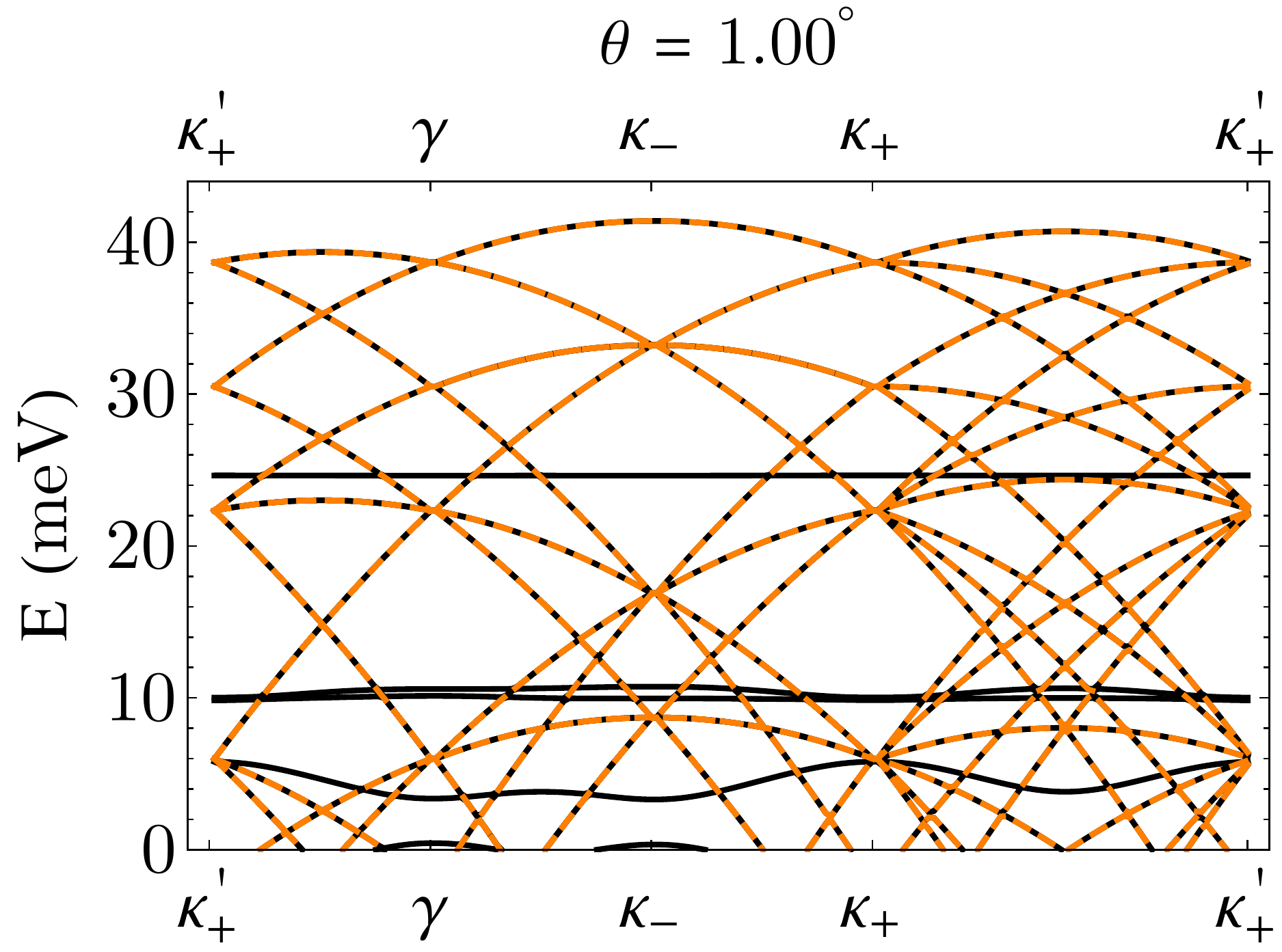}
\caption{Moir\'e band structures of $C_t^{-1}$ for twist angle $\theta=1.00^\circ$. The superimposed orange dashed lines refer to the band structure of the bottom right entry of the moir\'e Hamiltonian in Eq.\ \eqref{eq:moire-hamiltonian}.}\label{fig:Ct(-1)}
\end{figure}

\subsection{Top twist with a negative high-symmetry shift between the bottom two layers $C_t^{-1}$} 

Next, we investigate configuration, $C_t^{-1}$ (see Tab.\ \ref{tab:my_label}), which is a trilayer where the top layer is rotated with respect to the bottom two layers and there is a high-symmetry relative shift between the bottom two layers, i.e., the third layer is displaced by $-(\mathbf{a}_1+\mathbf{a}_2)/3$.

Fig.\ \ref{fig:Ct(-1)} shows the band structure of $C_{t}^{-1}$ for $\theta=1.00^\circ$. We observe the appearance of parabolic bands. This observation indicates that the trilayer structure effectively decouples into a bilayer and single-layer structure. That this is indeed the case can be seen from the moir\'e Hamiltonian in Eq.\ \eqref{eq:moire-hamiltonian}. Indeed, the interlayer coupling term $ \Delta^T_{23}(\boldsymbol{\delta}_3^{-1})$ vanishes for all angles in this case. To demonstrate this visually, we have superimposed an orange dashed line on top of the band structure in Fig.\ \ref{fig:Ct(-1)}. The orange line corresponds to the band structure of the bottom right entry of Eq.\ \eqref{eq:moire-hamiltonian} only. Note also the presence of various flat bands. We observe one at $E \approx 25$ meV and a pair around $E \approx 10$ meV. These bands arise from the decoupled bilayer layer structure, and they have valley Chern numbers, counting from above, of 0, $-1$, and 1, respectively.

\begin{figure*}[t!]
\centering\graphicspath{{./Figures/}}
\includegraphics[width=0.75\linewidth]{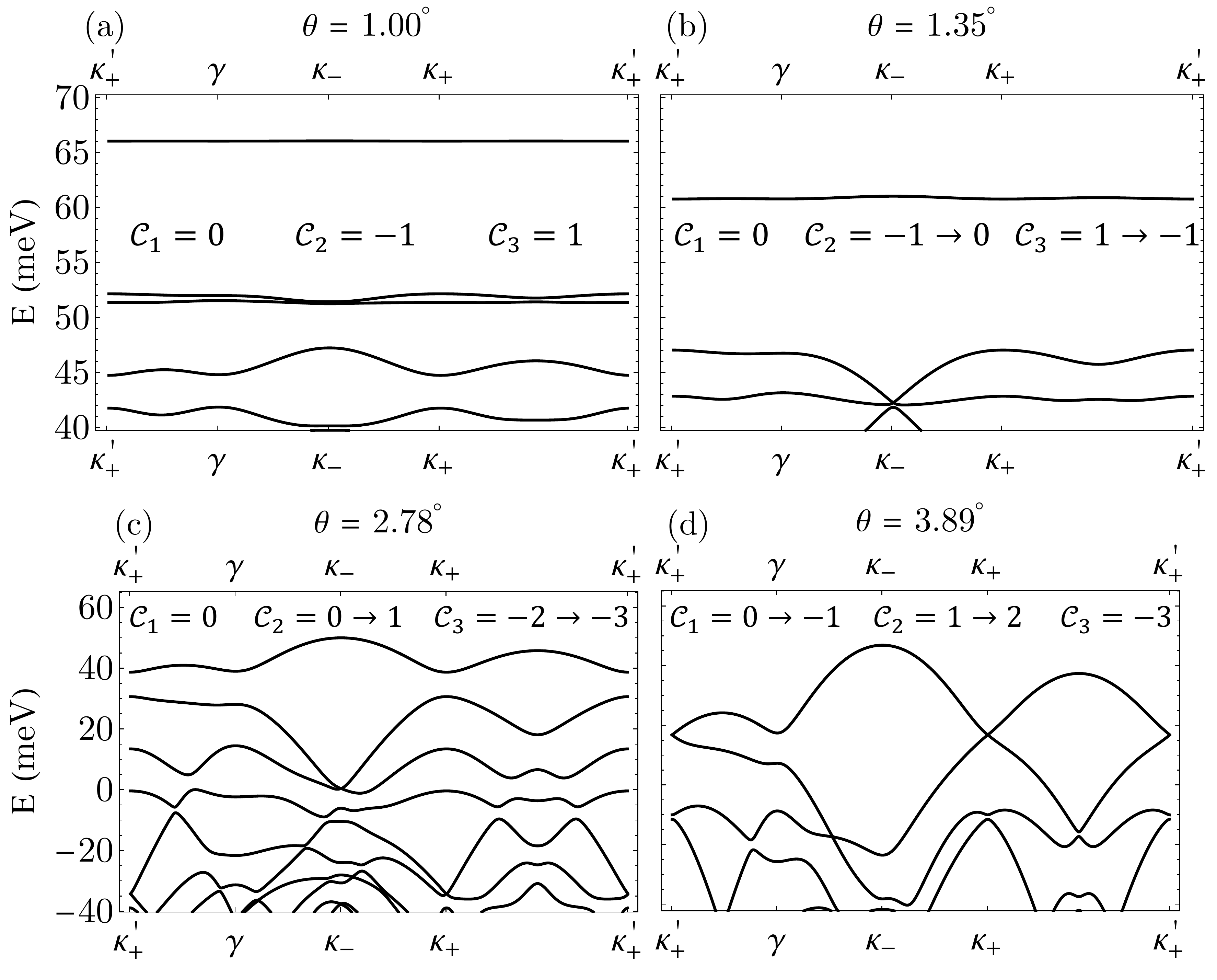}
\caption{Moir\'e band structures of configuration $C_t^1$ for twist angles (a) $\theta=1.00^\circ$, (b) $\theta=1.35^\circ$, (c) $\theta=2.78^\circ$, and (d) $\theta=3.89^\circ$. }\label{fig:Ct1}
\end{figure*}

\begin{figure*}[t!]
\centering\graphicspath{{./Figures/}}
\includegraphics[width=0.75\linewidth]{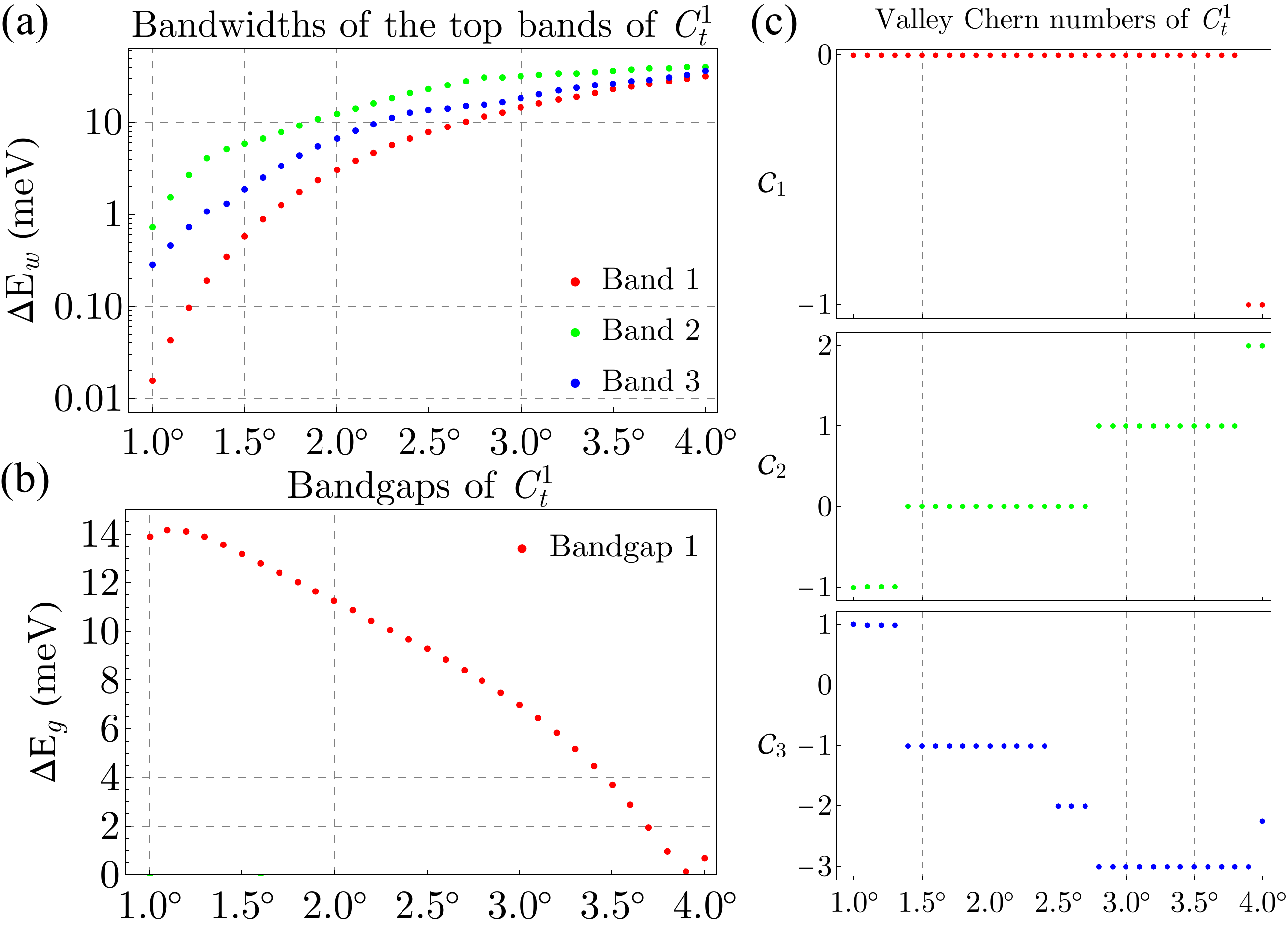}
\caption{Summary of $C_t^1$ calculations. (a) Bandwidths of the top three bands in a logarithmic scale of energy at different twist angles $\theta$. (b) The first two bandgaps as a function of $\theta$, where bandgap 1 refers to the gap between band 1 and band 2 (c) Valley Chern numbers of the top three bands as a function of $\theta$. Note that the gap between band 2 and band 3, referred to in other configurations as bandgap 2, is zero in this case for all twist angles.}\label{fig:Ct1-summary}
\end{figure*}

\subsection{Top twist with a positive high-symmetry shift between the bottom two layers $C_t^1$} 

As a third configuration we study, $C_t^1$ (see Tab.\ \ref{tab:my_label}). This is a trilayer with the top layer rotated relative to the bottom two layers and the third layer shifted by the high-symmetry displacement $(\mathbf{a}_1+\mathbf{a}_2)/3$ relative to the second layer. 

Again starting from a twist angle $\theta=1.00^\circ$ we see the band structure in Fig.\ \ref{fig:Ct1}(a). We observe that all three top most bands exhibit extreme flatness with $\Delta E_w<1$ meV. The top band at $E=66$ meV is exceptionally flat with a width of only $ \Delta E_{w1} \approx 0.015$ meV, however, it is topologically trivial. This is contrasted by the second and third bands, which are in close proximity to one another and located near $E \approx 52$ meV. As we see in Fig.\ \ref{fig:Ct1-summary}(b), they are topologically non-trivial with valley Chern numbers $\mathcal{C}_2=-1$ and $\mathcal{C}_3=1$. 

It is important to notice that the spacing between bands is larger than the top-twist case with no relative shift between the bottom two layers $C_t^0$. This is because the Hamiltonian decouples into a block for the top two layers and a block for the bottom two layers, similar to the case with a negative shift between the bottom two layers. However, unlike the previous case, the energies associate with the single-layer block are shifted down by approximately $\sim80$ meV, which is why we do not see parabolic bands in our plots. The same reasoning also explains why the bands shown in the figures bear a lot of similarity to the twisted homobilayer that was studied in Refs.\ \cite{Wu2019,Vogl_2021_floquet_homobilayerTMD} - differences arise from the $\Delta_{2b}$ term in Eq.\ \eqref{eq:moire-hamiltonian}.

From Fig.\ \ref{fig:Ct1-summary}(c), we see that the first topological phase transition happens somewhere between angles $\theta=1.3^\circ$ and $\theta=1.4^\circ$. However, unlike previously discussed transitions, the sum of valley Chern numbers on a first glance seems not to be conserved. However, this discrepancy is easily resolved when we consider a finer twist angle mesh. Indeed, when at twist angle $\theta=1.32^\circ$, the third band touches a lower band (band 4) and yields a topological transition $\mathcal{C}_3=1\xrightarrow{}0$. As we slightly increase the twist angle to $\theta=1.35^\circ$, the second and third bands touch at $\mathbf{k}=\boldsymbol\kappa_-$, which is accompanied by the following changes in valley chern numbers $\mathcal{C}_2=-1\xrightarrow{}0$ and $\mathcal{C}_3=0\xrightarrow{}-1$ (see Fig.\ \ref{fig:Ct1}(b)). At the same angle we also observe the first and third bands remain very flat. In fact the first band has a bandwidth $\Delta E_w<2$ meV for angle ranges up to $\theta \approx 1.8^\circ$, whereas the third band reaches the same bandwidth at $\theta \approx 1.5^\circ$. This has be contrasted with the previously studied configuration $C_t^0$ (see Sec.\ \ref{sec:toptwst_noshift}), where both the first and third bands reached bandwidths $\Delta E_w=2$ meV much earlier for a twist angle $\theta \approx 1.5^\circ$.      

The bands of the $C_{t}^1$ configuration undergo several additional topological transitions as the twist angle $\theta$ increases. In Fig.\ \ref{fig:Ct1-summary}(c), we see that the valley Chern number of the third band changes $\mathcal{C}_3=-1\xrightarrow{}-2$ as it touches a lower band (band 4) at and an angle $\theta=2.41^\circ$. At $\theta \approx 2.78^\circ$, the second band crosses the third band at $\mathbf{k}=\boldsymbol \kappa_-$, as shown in Fig.\ \ref{fig:Ct1}(c), which change their valley Chern numbers according to $\mathcal{C}_2=0\xrightarrow{}1$ and $\mathcal{C}_3=-2\xrightarrow{}-3$. Last, at $\theta=3.85^\circ$, the first band touches the second band at $\mathbf{k}=\boldsymbol \kappa_+$ (and also at $\mathbf{k}=\boldsymbol \kappa_+'$ due to inversion symmetry in k-space). The first band 
 now becomes topologically non-trivial for the first time because $\mathcal{C}_1=0\xrightarrow{}-1$ and $\mathcal{C}_2=1\xrightarrow{}2$ (see Fig.\ \ref{fig:Ct1}(d)). These two transitions show up as kinks in Fig.\ \ref{fig:Ct1-summary}(b). Finally, notice that at $\theta=4.0^\circ$, $\mathcal{C}_3$ is not an integer, since the third band is touching the 4th band. 

\begin{figure*}[t!]
\centering\graphicspath{{./Figures/}}
\includegraphics[width=0.75\linewidth]{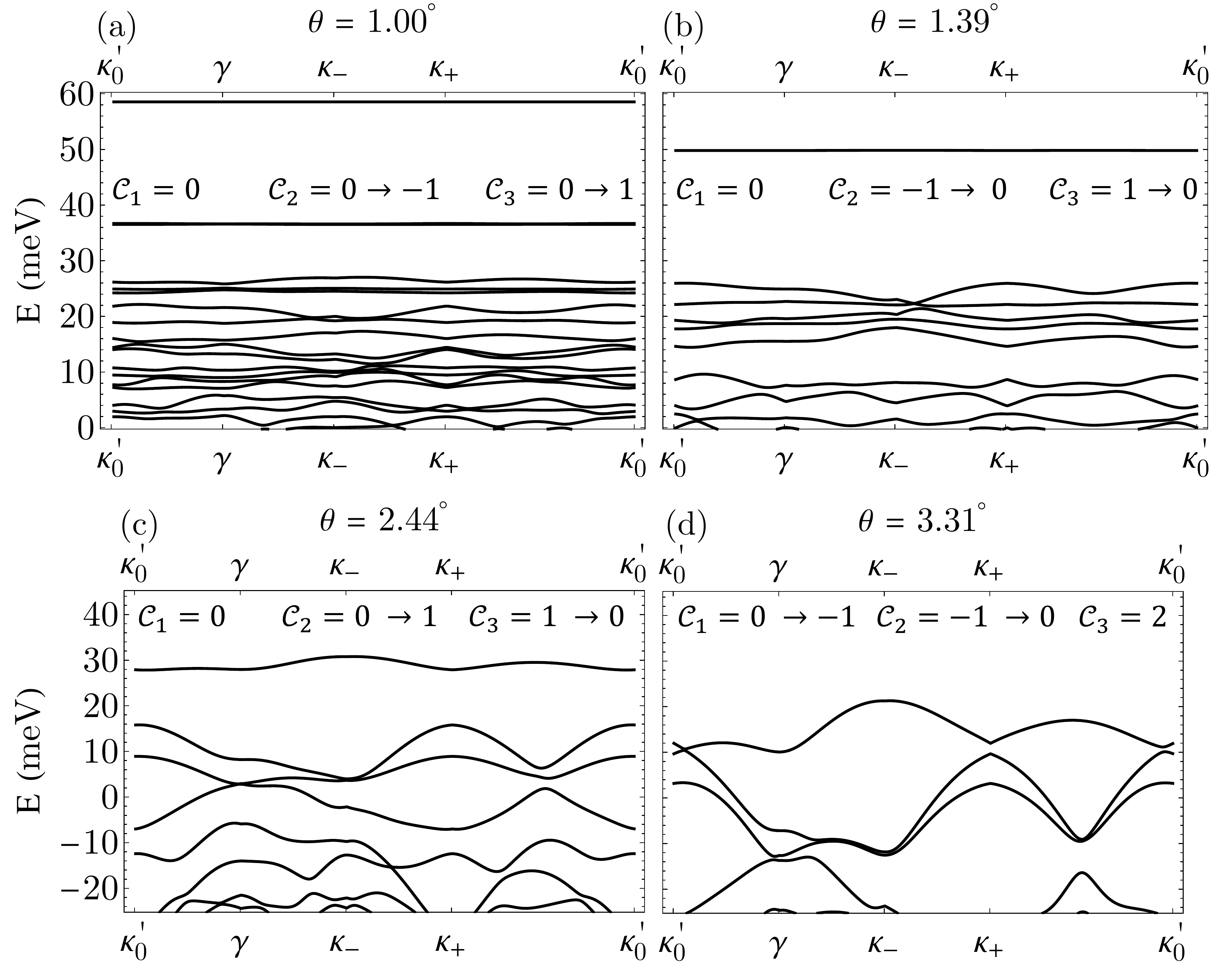}
\caption{Moir\'e band structures of $C_m^n$ for twist angles (a) $\theta=1.00^\circ$, (b) $\theta=1.39^\circ$, (c) $\theta=2.44^\circ$, and (d) $\theta=3.31^\circ$. Note that in this case we choose  a different high-symmetry path than before to capture the main features of the configurations. }\label{fig:Cm}
\end{figure*}

\begin{figure*}[t!]
\centering\graphicspath{{./Figures/}}
\includegraphics[width=0.75\linewidth]{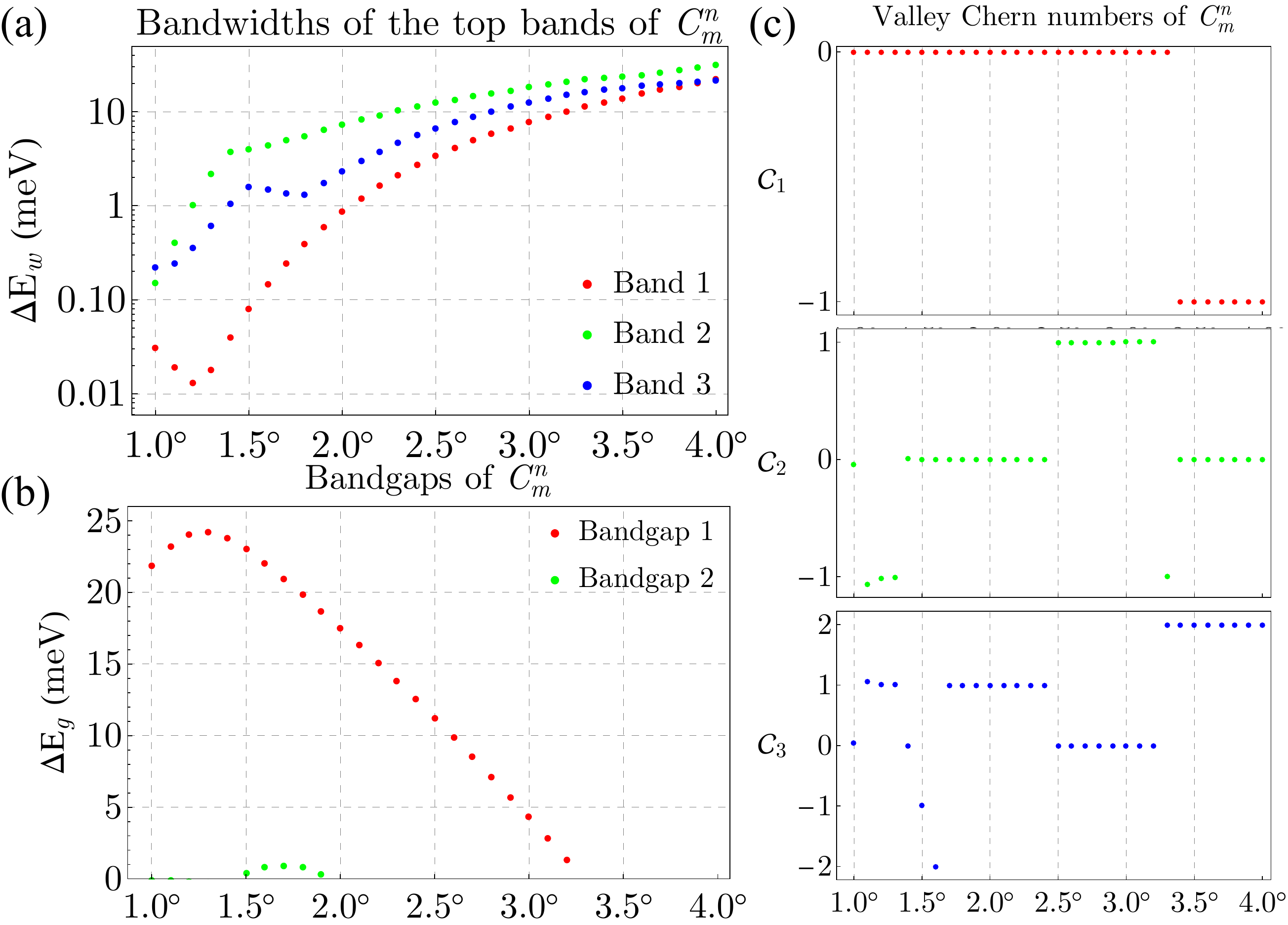}
\caption{Summary of $C_m^n$ calculations. (a) Bandwidths of the top three bands in a logarithmic scale of energy at different twist angles $\theta$. (b) The first two bandgaps as a function of $\theta$, where bandgap 1 refers to the gap between band 1 and band 2, and bandgap 2 refers to the gap between band 2 and band 3. (c) Valley Chern numbers of the top three bands as a function of $\theta$.}\label{fig:Cm-summary}
\end{figure*}

\subsection{Middle twist for any high-symmetry shift between the bottom two layers $C_m^n$} 
The last configurations we investigate are the trilayers where the middle layer is rotated with respect to the other two layers, (see the last three rows in Tab.\ \ref{tab:my_label}). Interestingly, our calculations show that all three configurations display identical band structures regardless of the high-symmetry displacement of layer 3, i.e., $C_m^0$, $C_m^{-1}$, and $C_m^1$ are equivalent for the purpose of this study. Hence, we refer to them collectively as $C_m^n$.

Fig.\ \ref{fig:Cm}(a) shows the band structures of $C_{m}^{n}$ for $\theta=1.00^\circ.$ Similar to the previous configuration $C_t^1$, all top bands are extremely flat, with the second and third bands reaching unprecedented flatness. The first band, at $E=59$ meV, is topologically trivial and it is separated by a large bandgap from the second and third bands, which appear as a single band in Fig.\ \ref{fig:Cm}(a) due to their extremely small separation (our calculations show that $\Delta E_{g2}=-0.06$ meV). At this angle, $\theta=1.00^\circ$, the latter two bands are on the verge of a topological transition that changes their valley Chern numbers according to $\mathcal{C}_2=0\xrightarrow{}-1$ and $\mathcal{C}_3=0\xrightarrow{}1$. 

We take a closer look at the bandwidths of the top three bands in Fig.\ \ref{fig:Cm-summary}(a). Interestingly, we observe that the first band reaches its minimum bandwidth at $\theta=1.2^\circ$, $\Delta E_w\approx 0.013$ meV, a behaviour that we did not see in the previous configurations in which the width of the first band always increased with $\theta$. In this configuration we also see that the width of band 2 is for the first time comparable in flatness to the other two bands at the first few twist angles. This means that we are observing an ultra flat and topologically non-trivial band. A width of $\Delta E_w=2$ meV (previously set arbitrarily to characterize flatness) is reached by the second band at $\theta \approx 1.3^\circ$, by the third band at $\theta \approx 2.0^\circ$, and by the first band at $\theta \approx 2.3^\circ$, which are significantly wider angle ranges than those displayed by the previous configurations we discussed, $C_t^0$ and $C_t^1$.

In addition to being superior to the previous configurations when it comes to the appearance of interesting flat bands, $C_m^n$ exhibits at $\theta=1.0^\circ$ a considerably larger bandgap of $\Delta E_{g1}=22$ meV between the first and second bands (see Fig.\ \ref{fig:Cm-summary}(b)). This bandgap reaches its maximum at $\theta=1.3^\circ$ with $\Delta E_{g1}=24$ meV. 

The second topological transition takes place at $\theta=1.39^\circ$, see Fig.\ \ref{fig:Cm}(b), when the second band and third band cross at a k-point between $\boldsymbol \kappa_-$ and $\boldsymbol \kappa_+$ to yield a change in valley Chern numbers$\mathcal{C}_2=-1\xrightarrow{}0$ and $\mathcal{C}_3=1\xrightarrow{}0$. Afterwards, the third band undergoes a series of topological transitions as it touches a lower band (band 4) multiple times, resulting in the following changes in its valley Chern number: $\mathcal{C}_3=0\xrightarrow{}-1\xrightarrow{}-2\xrightarrow{}1$ at the twist angles $\theta=1.4^\circ, 1.5^\circ, 1.6^\circ,$ and $\theta=1.7^\circ$, respectively (see Fig.\ \ref{fig:Cm-summary}(c)). Later, when $\theta \approx 2.44^\circ$, there is a crossing between the second and third band close to the $\boldsymbol{\kappa_-}$ point( see Fig.\ \ref{fig:Cm}(c)). The bands swap valley Chern numbers as $\mathcal{C}_2=0\xrightarrow{}1$ and $\mathcal{C}_3=1\xrightarrow{}0$.

Last, in Fig.\ \ref{fig:Cm-summary}(c) we see that there are two different topological transitions that happen in the vicinity of $\theta=3.3^\circ$. The first happens at $\theta=3.26^\circ$, when the the second and third band touch so that $\mathcal{C}_2=1\xrightarrow{}-1$ and $\mathcal{C}_3=0\xrightarrow{}2$. A second one happens at $\theta=3.31^\circ$ (see Fig.\ \ref{fig:Cm}(d)), when the first band comes in contact with the second band for the first time. This band touching happens along the $\boldsymbol \kappa_0'-\boldsymbol \gamma$ path and yields a change in Chern numbers $\mathcal{C}_1=0\xrightarrow{}-1$ and $\mathcal{C}_2=-1\xrightarrow{}0$.

%
%

\section{Conclusion}\label{sec:conclusion}

We have studied the band structure and band topology of various stacking configurations of twisted TMD homotrilayers for small twist angles using a continuum model with a nearest-layer-hopping approximation. Some of the interesting results we obtained were extremely flat bands of bandwidths $<1$ meV separated by wide bandgaps and bands with relatively high valley Chern numbers. These effects were present in a variety of stacking configurations. Interestingly, the configuration with a top-twist and non-zero high-symmetry shifts between the bottom two layers gave different phenomenology. Here, we observe that the trilayer structure decouples, for all twist angles, into a bilayer and single-layer structure. This decoupling manifests itself in a band structure that consists of parabolic bands - a signature of a single layer - and flat bands at small twist angles - due to the twisted bilayer TMD that appears on top of the decoupled single layer. 

Our work demonstrates that twisted trilayer TMDs are a platform that hosts much exciting physics. Studies on twisted trilayer TMDs are scarce compared, for example, to studies on trilayer graphene \cite{Chen2019,Chen2019a,Chen2020,PhysRevB.104.195429, Park2021, Zhou2021, Lin2022, Turkel2022, Kim2022}. Additional work in understanding TMD trilayers can be very fruitful. Indeed, numerous directions can be explored following this work. For example, our theoretical model can be a starting point for further investigations on the effects of introducing new parameters that give a more accurate model, e.g., next-nearest layer hopping. Additionally, since it is a simple continuum model, it allows studying additional effects, such as differences in potential between layers, which could arise, for instance, due to a substrate or an external electric field. Another interesting direction would be to investigate the effects of changing the stacked TMD materials and thus consider hetrotrilayer structures. 

Our study reveals aspects of the richness of trilayer TMDs compared to their bilayer counterparts. With the rapid development of experimental techniques, it can be expected that control over the stacking configurations of moir\'e materials will vastly improve in the future. It is possible that such improvements might lead to future applications of different trilayer TMD configurations in technology. After all, we have demonstrated that many physical properties could be highly tunable for intended applications.

\bibliography{references}
\end{document}